\documentclass[iop]{emulateapj}
\begin{document}

\title{Is WISEP J060738.65+242953.4 Really A Magnetically Active, Pole-on L Dwarf?}

\author{Matthew Route\altaffilmark{1}}

\altaffiltext{1}{Research Computing, Information Technology at Purdue, Purdue University, 155 S. Grant St., West Lafayette, IN 47907, USA; mroute@purdue.edu}

\slugcomment{Accepted for publication in ApJ: 2017 June 9}

\begin{abstract}

The interplay of rotation and manifested magnetic activity on ultracool dwarfs (UCDs) is of key importance for gathering clues as to the operation of the dynamos within these objects. A number of magnetized UCDs host kG-strength magnetic fields.  It was recently reported that the L8 dwarf WISEP J060738.65+242953.4 is a radio-emitting UCD that is likely observed pole-on, due to its lack of photometric variability and narrow spectral lines.  Follow-up radio observations at Arecibo Observatory, together with careful analysis of previously published details, however, suggest that the scientific and statistical significance of the radio and spectroscopic data have been overstated.  If the UCD is observed along its aligned spin/magnetic axis, the absence of observed H$\alpha$ activity may present challenges to the auroral model of UCD magnetism, although short or long-term cyclic magnetic activity may explain this behavior.  Monte Carlo simulations presented here suggest that the source probably rotates with $v~sin~i$=6-12 km s$^{-1}$, indicating that its inclination angle and rotational velocity are unexceptional and that its angular momentum has evolved as expected for brown dwarfs observed in $\sim$1 Myr-old clusters.  The discovery and verification of the most rapidly and slowest rotating brown dwarfs place valuable constraints on the angular momentum evolution and magnetic activity history of these objects.

\end{abstract}

\section{Introduction}

Solar and low-mass stars are born with rotation rates ranging from $\sim$1-10 d that evolve in a similar manner.  During the accreting, pre-main sequence phase (PMS; 1-5 Myr), the spin rate of solar-mass stars remains relatively unchanged, while lower mass stars spin up.  As stars approach the zero-age main sequence (ZAMS; 5-40 Myr) their contraction causes them to spin up, although interactions with the circumstellar disk reduce this effect.  Solar-mass stars that were initially more slowly rotating may have the effects of contraction further muted by core/envelope decoupling.  Afterwards, magnetized stellar winds cause the rotational periods of these stars to lengthen (Bouvier et al. 2014 and references therein).

Throughout these phases, magnetic fields play an important role in guiding stellar angular momentum evolution.  The angular momentum evolution of brown dwarfs is less drastic, however, with rare evidence for disk locking, and little angular momentum loss thereafter.  Spanning these two regimes are the ultracool dwarfs (UCDs), with spectral types $\gtrsim$M7, that feature fully convective interiors and exhibit similar patterns of magnetic activity.  Their X-ray and H$\alpha$ emission decline with effective temperature, while their radio luminosities remain roughly constant \citep{mcl12,rou16b}.  \citet{rod09} proposed that persistent differences in brown dwarf magnetic field topology were responsible for the link between greater photometric variability and slower rotation of young substellar objects in the $\sim$1 Myr Orion Nebula Cluster (ONC).  On the other hand, \citet{rou16} argued that many UCDs exhibit solar-like magnetic activity cycles that result in periodic changes in magnetic topology.  Thus, the discovery of rapidly and slowly rotating UCDs is important not only to unlock the secrets of their internal dynamos that give rise to large-scale, kG magnetic fields, but also to aid our understanding of their angular momentum evolution.

\citet{cas12} indentified WISEP J060738.65+242953.4 (J0607+24) during a search for sources detected in the \emph{Wide-field Infrared Survey Explorer (WISE)} catalog without corresponding nearby Two Micron All Sky Survey (2MASS) sources within 3'', thereby indicating their large proper motions.  They computed a $7.8_{-1.2}^{+1.4}$ pc distance to the object based on averaging the results of absolute magnitude versus spectral type relationships that utilized the 2MASS $J$, $H$, and $K_{S}$ \citep{loo08}, and SDSS $i$ and ${z}$ \citep{sch10} bands.  To determine the spectral type of J0607+24, \citet{cas12} compared the $i-z$ and $i-J$ colors to the median spectral type colors of L0-L8 dwarfs \citep{sch10}, concluding that the colors best matched an L8 dwarf, although T dwarf colors were not included in that study.  Comparison of J0607+24's $W1-W2$, $J-W2$, and $H-W2$ colors to those of L0-T9 UCDs detected by WISE, though, confirmed its L8 classification \citep{mai11}, with an inferred $T_{eff}=1460 \pm 90$ K \citep{loo08}.  Follow up near infrared spectroscopy conducted at the NASA Infrared Telescope Facility (IRTF) with SpeX demonstrated that J0607+24's spectrum was best fit with an L8 optical/ L9 near infrared spectral standard \citep{cas13}.  The study also presented a preliminary parallax measurement for the source as $139\pm2$ mas, yielding a revised distance estimate of $7.19_{-0.10}^{+0.11}$ pc - the value adopted to compute physical quantities throughout the remainder of this paper.

Optical spectroscopy of J0607+24 with the Gemini-North telescope and GMOS spectrograph on 2012 October 15 failed to detect any H$\alpha$ emission with a limiting equivalent width of $<$0.5\AA \citep{giz16}.  However, lithium absorption was detected, which along with the measurement of $L_{bol}/L_{\odot}=-4.66\pm0.02$ ($T_{eff}=1250$ K), constrained J0607+24's mass to be M$\leq$0.055M$_{\odot}$, with age $t\lesssim$2 Gyr.  Keck II NIRSPEC spectroscopic observations conducted on 2013 October 16 enabled a measurement of $v~sin~i<6\pm2$ km s$^{-1}$.  Follow-up radio observations conducted in the detached 4.5-5.5 and 6.6-7.6 GHz bands with the Very Large Array (VLA) on 2014 May 23, reportedly detected quiescent, non-varying emission at 6.05 GHz.  Photometric monitoring with \emph{Spitzer} IRAC on 2014 May 23-24 and \emph{Kepler} failed to detect any variability $>$0.2\% and $>$1.4\%, respectively.

To better understand the interplay between rotation and magnetic activity, I conducted observations of J0607+24 at Arecibo Observatory, as described in Section 2.  Section 3 revisits the results of pervious radio and H$\alpha$ observations, followed by reevaluating the question of J0607+24's orientation. Section 4 discusses the opportunities that low $v~sin~i$ UCDs, such as J0607+24, offer to enhance our understanding of stellar magnetism, dynamo activity, and angular momentum evolution, while Section 5 summarizes the results of this paper and prescribes future research directions.

\section{Observations and Results}

Observations of J0607+24 were conducted over two consecutive nights with the 305-m Arecibo radio telescope, from 2016 December 13 to 2016 December 15 (MJDs 57736.15489 to 57736.25737, and 57737.15569 to 57737.25818) under observing program A3161. The observational set up featured the C-band receiver at a center frequency of 4.75 GHz and Mock Spectrometer with a $\sim$1.1 GHz bandpass, as recently described in detail \citep{rou16b}.  The acquisition of 4.7 hrs of radio data, binned at 0.9 s resolution, more than triples the quantity of radio observations of this source.

No flaring radio emission was detected.  Although the theoretical 1$\sigma$ sensitivity is $\sim$0.15 mJy, the presence of radio frequency interference (RFI) raises the detection threshold significantly. Using the 3$\sigma$ standard deviation of the time series integrated across the cleanest 172 MHz sub-band centered at 4.47 GHz, I compute an upper limit of 1.450 mJy for bursting Stokes V radio emission.  The comparison of this upper limit with the original, marginal detection of the source by \citet{giz16} is illustrated in Figure 1.  I note that this experimental set up would only detect radio flares with rise and decay times on the order of a few minutes that have $>$10\% circular polarization.

\section{Discussion of Prior Observations}
\subsection{Analysis of Previous Radio Observations}

The discovery of radio flares from sources as late as the T6.5 dwarf 2MASS J10475385+212423 (J1047+21) \citep{rou12} indicates that a significantly larger fraction of the atmosphere of a warmer, L8 UCD should be suitably ionized to couple to a kG-strength magnetic field and create non-thermal radio emission.  However, several cautionary remarks should be noted regarding the quiescent radio emission from J0607+24 described in \citet{giz16}.  I note that the source was reported to be detected with a ``total flux density of $15.6\pm6.3$ $\mu$Jy at a mean frequency of 6.05 GHz'', yielding a ``3.5$\sigma$ source'' detection \citep{giz16}.  Although the significance of the radio detection appears to be overstated, perhaps this calculation represents a 1$\sigma$ value of 4.4 $\mu$Jy beam$^{-1}$, propagated from unstated $\sigma_{rms}$ values at 5.0 and 7.1 GHz center frequencies.

I also note that although J0607+24 was observed at frequencies above and below 6.05 GHz, it was not observed within a 1.1 GHz bandpass surrounding the so-called mean frequency itself.  This may be problematic since simultaneous radio emission at multiple frequencies requires the plasma to occupy a large radial extent, which may be unlikely.  Radio monitoring of the M9 dwarf TVLM 513-46546 demonstrated that luminous radio bursts at 8.44 GHz had no corresponding bursts at 4.88 GHz, and that bursts at 4.88 GHz and 8.44 GHz had different helicities, indicating that they operated in regions of differing magnetic field orientation \citep{hal07}.  Similarly, \citet{ber09} detected periodic pulses from the L0+L1.5 system 2MASSW 0746425+200032 AB at 4.86 GHz without corresponding emission at 8.46 GHz.  On the other hand, non-flaring radio emission has been detected nearly simultaneously from TVLM 513-46546 at frequencies as detached as 1.4 GHz and 8.4 GHz \citep{ost06}.  In light of these considerations, it is therefore more appropriate to describe the quiescent radio emission as marginally detected, with less significant flux densities of $16.5\pm7.6$ $\mu$Jy (2.2$\sigma$) and $15.6\pm10.7$ $\mu$Jy (1.5$\sigma$) at center frequencies of 5.0 and 7.1 GHz, respectively.  Furthermore, the low significance of radio emission in these two bands makes the meaningful measurement of a spectral index, $\alpha$, (where $S_{\nu}\propto \nu^{\alpha}$) doubtful for J0607+24.

\citet{giz16} also reported a limit on the detection of circular polarization, $f_{circ}\lesssim$90\%, which, unfortunately, is insufficient to discriminate among various coherent and incoherent emission mechanisms.  These include: a steady cadence of low-energy reconnection events of polarized radiation similar to the smooth electron cyclotron maser (ECM) emissions found at Jupiter and Uranus, depolarized ECM, gyrosynchrotron, or synchrotron radiation \citep{zar98,hal08}.  Persistent, non-flaring radio radiation is generally thought to be caused by mildly-relativistic gyrosynchrotron emission given by $B_{G}\approx57 \nu_{GHz} \gamma^{-2}_{min}$, where $f_{circ}\approx 3/\gamma_{min}$ \citep{ber02}.  A circular polarization fraction of $\lesssim$90\% yields a minimum Lorentz factor of the electron population of $\gamma_{min}\sim3$, and a magnetic field ranging from $B_{low}\sim$26 G to $B_{hi}\sim$40 G.  Under the assumption of a dipolar magnetic field described by $B=B_{0}(a/r)^{3}$, where $a$ is the scale size of the field, the radial extent of plasma that emits simultaneously from 4.5-7.6 GHz is given by $r_{low}=(B_{hi}/B_{low})^{(1/3)}~r_{hi}$. This yields an estimate of the radius of the plasma at lower magnetic field strength, $r_{low}=1.19~r_{hi}$.  For $r_{hi}\sim 1 R_{J}$, this corresponds to a radial length scale $L\sim$10$^{9}$ cm, which is similar in thickness to the $\sim2 \times$10$^{8}$ cm solar chromosphere \citep{hall08}.  While it is difficult to determine their altitude of formation, cm-wavelength solar radio emissions are thought to be generated between the upper chromosphere and corona at a wide range of altitudes.  This range begins at a much smaller radius ($\sim 3 \times 10^{-3} R_{\odot}$) than the marginally-detected, quiescent radio emission at J0607+24, and continues to $\gtrsim 0.3 R_{\odot}$ (, $\sim2 \times 10^{10}$ cm), the scale height of coronal loops \citep{ver81,def07,shi11}.  Another data point for comparison comes from Very Long Baseline Array (VLBA) observations of the fully-convective dM5.5e UV Ceti B, which found that its magnetic loops extend greater than 2.2 stellar radii ($\sim2.2 \times 10^{10}$ cm) \citep{ben98}.  In contrast with both the Sun and UV Ceti B, the corona of J0607+24 appears to be surprisingly compact.  The brightness temperature is computed from $T_{b} = 2 \times 10^{9}~F_{\nu,mJy}\nu^{-2}_{GHz} d^{2}_{pc}(R/R_{J})^{-2}$ K, where $F_{\nu}$ is the measured radio flux density at emission frequency $\nu$, for a source at distance $d$.  If the emission spans the entire disk ($R=1.15 R_{J}$), $T_{b}\sim 2 \times 10^{7}$ K, which is consistent with gyrosynchrotron emission.

On the other hand, the maximum frequency of ECM-generated radio emission is the local cyclotron frequency, $\nu_{c}$, related to the local magnetic field strength by $\nu_{c} = 2.8 \times 10^{6} B_{G}$.  The VLA frequency range corresponds to $B_{hi}\leq2.7$ kG and $B_{low}\leq1.6$ kG, which yield $r_{low}=1.19~r_{hi}$.  Assuming the emission is confined to a polar cap at $\gtrsim$65$^{\circ}$ latitude \citep{bha00} that is viewed pole-on, and $R\sim 1~R_{J}$, the brightness temperature would be $T_{b}\gtrsim 9 \times 10^{8}$ K, consistent with ECM.  However, a persistent plasma cavity of $L\sim10^{9}$ cm has no precedent among better understood systems; although coronal shear flows above solar active regions start with $L\sim10^{7}-10^{9}$ cm, numerical simulations indicate that they readily decay into smaller structures.  Terrestrial cavities that create the auroral kilometric radiation measure only $L\sim10^{6}-10^{7}$ cm \citep{tre06}.

\subsection{Previous Spectroscopic Observations}

\citet{giz16} reported no H$\alpha$ detected from J0607+24 in either absorption or emission during 40 mins of observations.  It is tempting to conclude that this source does not emit H$\alpha$ and lacks magnetic activity.  In the auroral model of UCD magnetic activity, H$\alpha$ and radio emission are correlated \citep{hal15}and occur once per rotation.  Similarly, in the active region model, chromospheric H$\alpha$ emission is increased around active regions and network, and flares may occur more than once per rotation \citep{hall08}.  While this observation was long enough to capture one flare-per-rotation activity from ultra-short period UCDs, such as the T6-dwarf WISEPC J112254.73+255021.5 (J1122+25; Route \& Wolszczan 2016a) that potentially rotates as quickly as once every 0.288 hr, for other short period sources (e.g. 2MASS J22282889-4310262, $P$=1.41 hr) \citep{cla08}, it represents monitoring less than half a period.  Moreover, many UCDs would be observed for even smaller fractions of their rotation, since the UCD rotation period distribution can be approximated by a lognormal distribution with $\mu=$1.22 and $\sigma=$0.49 (hr) \citep{rad14,rou17}.  Thus, these observations cannot conclusively rule out H$\alpha$ emission that may result from either auroral or active region models of UCD magnetic reconnection activity.

\subsection{J0607+24 Orientation Considerations}

To examine the question of J0607+24's orientation, \citet{giz16} simulated $v~sin~i$ values using an isotropic distribution of inclination angles, $i$, coupled with rotational velocities derived from a lognormal rotational period distribution \citep{rad14}.  They found that $\sim3$\% of sources would have $v~sin~i<6$ km s$^{-1}$, and 77\% of these sources were observed roughly pole-on, defined as $i<20^{\circ}$.  The authors point out that this value may be suspect since the line broadening profile is smaller than the instrumental uncertainty, $\delta\nu=15$ km s$^{-1}$.

To better understand the effects of measurement uncertainty on this conclusion, I examine the influence of a measured $v~sin~i=$12 km s$^{-1}$ ($P\sim$10.2 hr).  This value represents the 3$\sigma$ formal uncertainty upper limit of the projected rotational velocity that \citet{giz16} found, and is somewhat smaller than the instrumental profile of the spectrograph.  10$^{7}$ simulations are executed that also assume an isotropic distribution of inclination angles that modify the lognormal rotational distribution parameterized by $\mu=1.22$ and $\sigma=0.49$, which \citet{rou17} found fits the UCD rotation periods measured by radio and photometric monitoring. Under these conditions, 10.2\% of UCDs would have $v~sin~i\leq$12 km s$^{-1}$, and only 45.6\% would be ``pole-on'' rotators.  Similarly, if we conduct 10$^{7}$ simulations using the original UCD lognormal rotational distribution of \citet{rad14}, described by the parameters $\mu=1.41$ and $\sigma=0.48$ (hr), 15.1\% of UCDs would have $v~sin~i\leq$12 km s$^{-1}$, and for only 34.7\% of objects would observers exclusively view a polar cap, meaning that $i>20^{\circ}$ is a much more likely orientation.  Thus, the probability of viewing J0607+24 pole-on strongly depends on $v~sin~i$: an increase of $v~sin~i$ from 6-12 km s$^{-1}$, which is reasonable due to formal uncertainty calculation and instrumental effects, changes the probable orientation of J0607+24 from pole-on and rapidly rotating to more inclined and slowly rotating.  This conclusion is not without precedent; as \citet{met15} says: \emph{``approximately a third of the periodicities are on timescales of $>$10 hr, suggesting that slowly rotating brown dwarfs may be common.''}

\citet{giz16} argued that since J0607+24 is observed along its rotation axis, radio-flaring activity is not expected.  Such flares are thought to be caused by ECM because of their $\sim$100\% circular polarization fractions and $\gtrsim$10$^{9}$ K brightness temperatures \citep{hal06}.  Models indicate that ECM radiation is comprised of RX-mode radiation emitted perpendicular to the local magnetic field with right-hand circular polarization, and LO-mode radiation emitted parallel to the field with left-hand circular polarization \citep{tre06}.  In many astrophysical contexts, RX-mode radiation appears to dominate, including at Jupiter, Saturn, and Uranus \citep{zar98} and as \citet{hal06} argued, for the M9 UCD TVLM 513-46546.  The failure to detect presumably RX-mode radio flares with Arecibo Observatory appears to support the hypothesis that J0607+24 is observed along its aligned spin/magnetic axis.  However, previous observations of radio-active sources highlight the sporadic nature of their flares and their large range of energies, thus making it difficult to draw firm conclusions \citep{rou16a,rou17}.  This reasoning also overlooks the detectability of weaker LO-mode radiation, which appears to cause various Jovian quasi-periodic bursts detected by \emph{Ulysses} \citep{zar98}.

The theoretical analysis of ECM continues to improve. \citet{mel99} argued that LO-mode emission was one means to avoid ECM radiative self-absorption by thermal electrons in the context of stellar radio emissions.  More recently, \citet{lee13} modelled loss cone electron distribution functions as a function of plasma-to-electron cyclotron frequencies ($\omega_{p}/\Omega$).  They found that the LO-mode dominates as harmonic modes are approached, resulting in emission angles corresponding to maximum growth, $\theta_{max}<15^{\circ}$, that are therefore nearly aligned with the magnetic field.  \citet{reg15} simulated the plasma emission properties of the solar corona above active regions using force-free models of the magnetic field together with hydrostatic models of the plasma.  Similarly, he found that while the RX-mode dominates for $\omega_{p}/\Omega\leq0.35$, LO-mode dominates for $0.35\leq\omega_{p}/\Omega\leq 1$.  These results suggest that the dominant emission mode depends on the altitude of the emitting plasma above the active regions, with smaller $\omega_{p}/\Omega$ values corresponding to stronger magnetic fields lower in the corona.  Thus, the notion that observing J0607+24 along its magnetic axis precludes the detection of polarized radio bursts is not supported on observational or theoretical grounds.

\section{Implications of J0607+24's Quiescent Radio Emission and Low Projected Rotational Velocity}
\subsection{The Inverse Relationship between Radio and X-ray Activity}

\citet{ste12} suggested that UCDs may be divided into two groups based on their observed radio and X-ray emission.  The first group exhibits flaring or steady X-ray emission, but does not emit radio flares, and includes sources such as LP412-31 and LHS 2065.  The second group, which includes objects such as TVLM 513-46546 and LSR J1835+3259, emit bright radio flares but X-ray emission is weak or absent.  Since only quiescent radio emission has been marginally detected from J0607+24 and no flaring radio emission has been detected, this classification scheme would lead us to expect that the source should have strong, potentially flaring, X-ray emission.

During a study of the X-ray luminosity fraction ($L_{X}/L_{bol}$) of 3 M6.5 dwarfs and 35 UCDs with measured $v~sin~i$ values, \citet{coo14} found that this emission measurement decreases with increasing rotational velocity.  Such a relationship may have its origins in the operation of a different dynamo in UCDs, centrifugal stripping of the coronal envelope, magnetic field topological effects, or for reasons still unknown.  Given J0607+24's low $v~sin~i$ of 6-12 km s$^{-1}$, this line of reasoning would also indicate that it should be a bright X-ray source.

Alternatively, the failure of observing campaigns to detect X-ray emission from sources cooler than the L2+L3.5 Kelu-1 AB system (T$_{eff}\sim$2000 K and 1800 K) \citep{aud07,coo14} may indicate that X-ray emission is strongly suppressed by the increasingly neutral atmospheres of these UCDs \citep{moh02}.  Recently, \citet{rod15} used DRIFT-PHOENIX to simulate dust grain formation, gas-dust interaction, and radiative transfer in 1D model atmospheres.  They characterized the decline in thermal ionization with later spectral types, and found that in the presence of a $\sim$kG magnetic field, the volume of magnetized atmosphere declines by roughly half an order of magnitude from 3000 K to 1000 K.  However, their simulations also indicated that a strong magnetic field causes plasma magnetization in the rarefied upper layers to a degree that is largely independent of T$_{eff}$.  Thus, magnetic reconnective activity such as X-ray and H$\alpha$ emission still occur on cooler UCDs, but with reduced emitting volumes.

The quiescent X-ray emission from the non-interacting binary L dwarf Kelu-1 AB system may be analogous to the X-ray emission from the L8 J0607+24.  Kelu-1 AB, located at a distance of 18.7 pc, was detected with \emph{Chandra X-ray Observatory}(CXO) by \citet{aud07}, who measured its 0.1-10 keV X-ray luminosity to be $L_{X}=2.9 \times 10^{25}$ erg s$^{-1}$ with a 23.8 ks total exposure time.  Since J0607+24 is only 7.8 pc away, if it has a quiescent X-ray luminosity similar to Kelu-1 AB, the object should be more readily detectable with CXO, as its closer proximity offers a factor of $\sim$6$\times$ improvement in X-ray flux.

\subsection{J0607+24 in the Activity Cycle Hypothesis}

H$\alpha$ observations conducted in 2012 October, and follow up radio and photometric monitoring in 2014 May failed to detect any significant magnetic activity.  Similarly, no activity was observed by Arecibo in 2016 December.  One potential conclusion from these sparse, and typically short duration, observations is that J0607+24 is magnetically active, but does not appear so.  This could be a result of the sporadic nature of the radio flares (e.g. J1122+25) \citep{rou16a}.  Yet even long-term \emph{Spitzer} and \emph{Kepler} observations that spanned $\sim$2 and $>$84 10.2-hr rotations, respectively, failed to detect magnetized spots that may rotate in and out of view.  \citet{met15} estimated that if spots occur with equal probability across the surface of a given brown dwarf, only $\sim$22\% of objects should have undetectable variability.

\citet{hal15} suggested that magnetized activity would not occur randomly across a UCD surface, but would be concentrated toward the magnetic poles of a strong, global, dipolar magnetic field.  The model for such activity are the radio - X-ray auroral emissions from Jupiter, which appear to be spatially correlated and clustered around the poles, pointing to their common origin in electron precipitation from the solar wind \citep{zar98,bha00}.  Similarly, \citet{hal15} argued that the simultaneous radio and H$\alpha$ emissions from the M8.5 dwarf LSR J1835+3259 were caused by reconnection events that accelerate electrons toward the atmosphere of the UCD at the poles.  According to this model, if we view J0607+24 along its aligned spin/magnetic axis, auroral ``hot spots'' and their luminous emissions should always be in view.  As these are not observed, but a low $v~sin~i$ is measured, perhaps the magnetic and spin axes are misaligned. The obvious analogy for this is Uranus, which radiates both steady and bursty broadband (b-smooth and b-bursty) radio emission components that are potentially ECM-induced \citep{zar98}.  Although marginally-significant radio emission has been detected, the short and non-simultaneous H$\alpha$ observations prevent the evaluation of these scenarios.

Alternatively, \citet{rou16} suggested that UCDs experience magnetic activity cycles similar to those found on the Sun and potentially M dwarfs as well \citep{rob13}.  In this picture, magnetic activity concentrates from mid-latitudes to the equator during the ``solar maximum'' phase when a weaker dipolar and stronger toroidal field exists, while the dipolar field strengthens during the ``solar minimum'' phase of the cycle.  If such UCD activity cycles exist, J0607+24 may have an activity cycle that is $>$8 years, which is double the time the source has been monitored without the detection of H$\alpha$ emission or radio bursts.  By comparison, \citet{rob13} found that $\gtrsim$14\% of their sample of M0-M5 stars may have activity cycles $>$1 yr in duration, and 7.5\% show evidence of magnetic activity cycles $>$4 years.  The Sun also experiences even longer-term cyclic changes in global magnetic activity, such as the Gleissberg Cycle and the secular trend.  The most familiar result of these longer cycles was a 70-yr period of extremely low levels of activity and the absence of sunspots, called the Maunder Minimum \citep{hat10}.  Similarly, \citet{ola09} found evidence for multidecadal activity trends in $\sim$55\% of G0-M2 stars.  While J0607+24's lack of activity other than marginal, comparatively weak, quiescent radio emission (Figure 1) may indicate that it hosts long-duration magnetic activity cycles, perhaps even including a Maunder Minimum phase, a $\sim$8-yr cycle would be rather typical compared to the durations of activity cycles measured on other stars.  More extensive monitoring of J0607+24 at multiple wavelengths would be required to evaluate these speculative scenarios.

The existence of activity cycles among brown dwarfs suggests that they do not have persistent magnetic topologies, potentially preventing topological effects from causing differences in their spin down rates  at $<$5 Myr (cf. \citet{rod09}).  0.4-1.2 M$_{\odot}$ stars are also born at a range of rotational periods, yet it has not been proposed that this variation is caused by topological differences, presumably on account of their well-known magnetic activity cycles.  Furthermore, recent theoretical work indicates that the efficiency with which large-scale dipolar fields cause angular momentum loss among solar-mass stars is significantly reduced, thereby diminishing the differences among topological effects.  Instead, this would suggest that among brown dwarfs as among solar-mass stars, disparities in their rotation rates are most likely related to the timescale for protostellar disk dispersion \citep{bou14}.

\subsection{The Significance of Low $v~sin~i$ for UCD Angular Momentum Evolution Models}

Only 5 UCDs have known rotational periods $P\gtrsim$10.2 hr ($v~sin~i\lesssim$12 km s$^{-1}$), including the L5.5+T5 system SDSS J151114.66+060742.9 AB, the L6 dwarf 2MASS J21481628+4003593, J0607+24, the L8+T3.5 system 2MASS J13243559+6358284 AB, and the T2.5 dwarf HN PegB- most discovered by the \emph{Spitzer} Weather on Other Worlds photometric observing campaign \citep{met15}.  Another two \emph{Spitzer}-monitored objects, the L4 dwarf 2MASS J16154255+4953211 (J1615+49) and the T0.5 dwarf SDSS J151643.01+305344.4 (J1516+30), may have rotational periods that exceed their observation lengths, at 24-hr and 50-hr, respectively.  Alternatively, these latter two objects may rotate more quickly, yet simply have little photometric variability.

Extremely rapidly and slowly rotating UCDs with known rotational periods have the potential to impose valuable constraints on the angular momentum evolution of these objects.  The recent detection of J1122+25, a powerful radio-flaring brown dwarf with a rotational period potentially as rapid as 0.288 hr, probes the lower limit of angular momentum loss by brown dwarfs.  This object could represent an instance where almost no angular momentum loss occurred.  Similarly, the five UCDs mentioned above explore the maximum extent of UCD spin-down.  Figure 8 in \citet{bou14} displays how brown dwarfs rotate approximately an order of magnitude faster as they age from 1 Myr to 1 Gyr.  Brown dwarf rotation rates increase due to the combined effects of gravitational contraction and the $\sim$10$^{4}$ reduced efficiency of rotational braking by magnetized winds compared to solar-type stars.  A similar angular momentum evolutionary path for J0607+24 currently rotating at $v\sim$6 km s$^{-1}$ implies a rotational period of $\sim$8 days at an age of 1 Myr.  This inferred rotational period for a young J0607+24 would be unexceptional, since photometric monitoring of brown dwarfs in the ONC revealed rotational periods $\leq$13 d \citep{rod09}.  However, this analysis indicates that verifying the rotational periods of J1615+49 and J1516+30 could place important constraints on the upper limits of spin down among brown dwarfs.

\section{Conclusion}

The failure of Arecibo to detect radio flaring emission from J0607+24 may have many causes, including the sporadic nature of flares, their varying energetics, geometric effects, and the influence of magnetic topology and activity cycles.  This work has reevaluated previous observations of magnetic activity indicators and found that the arguments for pole-on rotation and magnetic activity are rather less compelling that indicated.  This paper highlights the need for more rigorous experimental design to better constrain and examine the various magnetic phenomena that UCDs exhibit.  For UCDs that lack apparent photometric variability, radio monitoring may provide the best method to determine their rotational periods, thereby allowing inclination angle effects to be ignored.  Finally, evaluation of the potential existence of magnetic activity cycles in brown dwarfs is key to understanding whether different magnetic field topologies cause differing angular momentum evolutionary tracks, as proposed by \citet{rod09}.
 
\section{Acknowledgements}

M.R. acknowledges publication support from the Purdue APSAC Professional Development grant.  The Arecibo Observatory is operated by SRI International under a cooperative agreement with the National Science Foundation (AST-1100968), and in alliance with Ana G. M\'{e}ndez-Universidad Metropolitana, and the Universities Space Research Association.

\clearpage

\begin{figure}
\centering
\includegraphics[width=1.1\textwidth,angle=0]{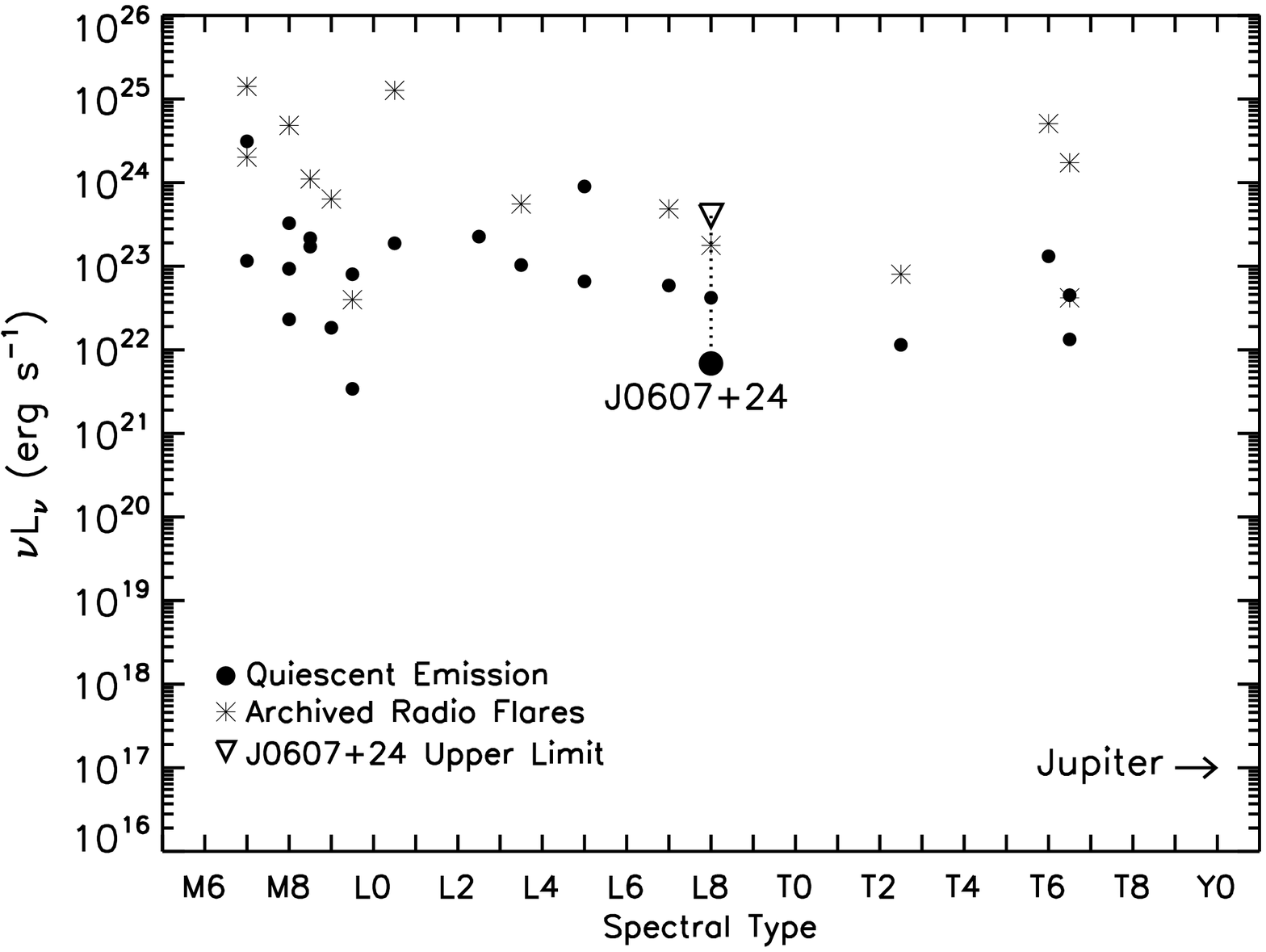}
\caption{Logarithmic plot of the radio luminosity, $\nu$L$_{\nu}$, versus spectral type of detected radio-loud UCDs (\citet{wil14,bur15,rou16b} and references therein). The large filled circle denotes the marginal detection of quiescent radio emission from J0607+24 \citep{giz16}, while the upper limit on flaring activity constrained by Arecibo is denoted by the inverted triangle above it.  On account of J0607+24's proximity to the Sun, both measurements are quite low compared to other UCDs.  For comparison, Jupiter's radio emission \citep{laz07} places it off to the lower right of the diagram.  \label{fig1}}
\end{figure}


\clearpage


\begin{thebibliography}

\bibitem[Audard et al.(2007)]{aud07}Audard, M., Osten, R. A., Brown, A., Briggs, K. R., G\"{u}del, M., et al. 2007, A\&A, 471, L63
\bibitem[Benz et al.(1998)]{ben98}Benz, A. O., Conway, J., \& G\"{u}del, M. 1998, A\&A, 331, 596
\bibitem[Berger(2002)]{ber02}Berger, E. 2002, ApJ, 572, 503
\bibitem[Berger et al.(2009)]{ber09}Berger, E., Rutledge, R. E., Phan-Bao, N., et al. 2009, ApJ, 695, 310 2009, ApJ, 695, 310
\bibitem[Bhardwaj \& Gladstone(2000)]{bha00}Bhardwaj, A., \& Gladstone, G. R. 2000, Rev. Geophys., 38, 3, 295
\bibitem[Bouvier et al.(2014)]{bou14}Bouvier, J., Matt, S. P., Mohanty, S., et al. 2014, in Protostars and Planets VI, ed. H. Beuther et al. (Tucson, AZ: Univ. Arizona Press), 43
\bibitem[Burgasser et al.(2015)]{bur15}Burgasser, A. J., Melis, C., Todd, J., Gelino, C. R., Hallinan, G., et al. 2015, AJ, 150, 180
\bibitem[Castro \& Gizis(2012)]{cas12}Castro, P. J., \& Gizis, J. E. 2012, ApJ, 746, 3
\bibitem[Castro et al.(2013)]{cas13}Castro, P. J., Gizis, J. E., Harris, H. C., Mace, G. N., Kirkpatrick, J. D., et al. 2013, ApJ, 776, 126
\bibitem[Clarke et al.(2008)]{cla08}Clarke, F. J., Hodgkin, S. T., Oppenheimer, B. R., Robertson, J., \& Haubois, X. 2008, MNRAS, 386, 2009
\bibitem[Cook et al.(2014)]{coo14}Cook, B. A., Williams, P. K. G., \& Berger, E. 2014, ApJ, 785, 10
\bibitem[DeForest(2007)]{def07}DeForest, C. E. 2007, ApJ, 661, 532
\bibitem[Gizis et al.(2016)]{giz16}Gizis, J. E., Williams, P. K. G., Burgasser, A. J., Libralato, M., Nardiello, D., et al. 2016, AJ, 152, 123
\bibitem[Hall(2008)]{hall08}Hall, J. C. 2008, Living Rev. Solar Phys., 5, 2
\bibitem[Hallinan et al.(2006)]{hal06}Hallinan, G., Antonova, A., Doyle, J. G., et al. 2006, ApJ, 653, 690
\bibitem[Hallinan et al.(2007)]{hal07}Hallinan, G., Bourke, S., Lane, C., et al. 2007, ApJL, 663, L25
\bibitem[Hallinan et al.(2008)]{hal08}Hallinan, G., Antonova, A., Doyle, J. G., et al. 2008, ApJ, 684, 644
\bibitem[Hallinan et al.(2015)]{hal15}Hallinan, G., Littlefair, S. P., Cotter, G., et al. 2015, Natur, 523, 568
\bibitem[Hathaway(2010)]{hat10}Hathaway, D. H. 2010, Living Rev. Solar Phys., 7, 1
\bibitem[Lazio \& Farrell(2007)]{laz07}Lazio, T. J. W., \& Farrell, W. M. 2007, ApJ, 668, 1182
\bibitem[Lee et al.(2013)]{lee13}Lee, S.-Y., Yi, S., Lim, D., et al. 2013, JGRA, 118, 7036
\bibitem[Looper et al.(2008)]{loo08}Looper, D. L., Gelino, C. R., Burgasser, A. J., \& Kirkpatrick, J. D. 2008, ApJ, 685, 1183
\bibitem[Mainzer et al.(2011)]{mai11}Mainzer, A., Cushing, M. C., Skrutskie, M., et al. 2011, ApJ, 726, 30
\bibitem[McLean et al.(2012)]{mcl12}McLean, M., Berger, E., \& Reiners, A. 2012, ApJ, 746, 23
\bibitem[Melrose(1999)]{mel99}Melrose, D. B. 1999, Astrophys. Space Sci., 264, 391
\bibitem[Metchev et al.(2015)]{met15}Metchev, S. A., Heinze, A., Apai, D., Flateau, D., Radigan, J., et al. 2015, ApJ, 799, 154
\bibitem[Mohanty et al.(2002)]{moh02}Mohanty, S., Basri, G., Shu, F., \& Allard, F. 2002, ApJ, 571, 469
\bibitem[Ol\'{a}h et al.(2009)]{ola09}Ol\'{a}h, K., Koll\'{a}th, Z., Granzer, T., Strassmeier, K. G., Lanza, A. F., et al. 2009, A\&A, 501, 703
\bibitem[Osten et al.(2006)]{ost06}Osten, R. A., Hawley, S. L., Bastian, T. S., \& Reid, I. N. 2006, ApJ, 637, 518
\bibitem[Radigan et al.(2014)]{rad14}Radigan, J., Lafreni\'{e}re, D., Jayawardhana, R., Artigau, E. 2014, ApJ, 793, 75
\bibitem[R\'{e}gnier(2015)]{reg15}R\'{e}gnier, S. 2015, A\&A, 581, A9
\bibitem[Robertson et al.(2013)]{rob13}Robertson, P., Endl, M., Cochran, W. D., \& Dodson-Robertson, S. E. 2013, ApJ, 764, 3
\bibitem[Rodr\'{i}guez-Barrera et al.(2015)]{rod15}Rodr\'{i}guez-Barrera, M. I., Helling, C., Stark, C. R., \& Rice, A. M. 2015, MNRAS, 454, 3977
\bibitem[Rodr\'{i}guez-Ledesma et al.(2009)]{rod09}Rodr\'{i}guez-Ledesma, M. V., Mundt, R., \& Eisl\"{o}ffel, J. 2009, A\&A, 502, 883
\bibitem[Route(2016)]{rou16}Route, M. 2016, ApJL, 830, L27
\bibitem[Route(2017)]{rou17}Route, M. 2017, \emph{submitted to ApJ}
\bibitem[Route \& Wolszczan(2012)]{rou12}Route, M. \& Wolszczan, A. 2012, ApJL, 747, L22
\bibitem[Route \& Wolszczan(2016a)]{rou16a}Route, M. \& Wolszczan, A. 2016a, ApJL, 821, L21
\bibitem[Route \& Wolszczan(2016b)]{rou16b}Route, M. \& Wolszczan, A. 2016b, ApJ, 830, 85
\bibitem[Schmidt et al.(2010)]{sch10}Schmidt, S. J., West, A. A., Hawley, S. L., \& Pineda, J. S. 2010, AJ, 139, 1808
\bibitem[Shibasaki(2011)]{shi11}Shibasaki, K., Alissandrakis, C. E., \& Pohjolainen, S. 2011, SoPh, 273, 309
\bibitem[Stelzer et al.(2012)]{ste12}Stelzer, B., Alcal\'{a}, J., Biazzo, K., Ercolano, B., Crespo-Chac\'{o}n, I., et al. 2012, A\&A, 537, A94
\bibitem[Treumann(2006)]{tre06}Treumann, R. A. 2006, A\&AR, 13, 229
\bibitem[Vernazza et al.(1981)]{ver81}Vernazza, J. E., Avrett, E. H., \& Loeser, R. 1981, ApJS, 45, 635
\bibitem[Williams et al.(2014)]{wil14}Williams, P. K. G., Cook, B. A., \& Berger, E. 2014, ApJ, 785, 9
\bibitem[Zarka(1998)]{zar98}Zarka, P. 1998, J. Geophys. Res., 103, 20, 159

\end{thebibliography}
\end{document}